# Outcome-Classified Precision Auditing of Filter Rules in Algorithmic DEX Trading: Evidence from 2,400 Rejection Events

**Arati Uday Kamat**
Independent Researcher · ORCID: 0009-0000-4781-312X

## Abstract

This paper reports a precision audit of a production filter stack against a 13-day window of post-rejection forward-market observations on Solana DEX trading (2026-04-10 to 2026-04-23, UTC). The audit yielded 99,510 follow-up samples across 2,402 unique rejection events spanning eight active filter rules. We classify each event under a five-tier outcome rule and report per-filter distributions. The descriptive result is a save-to-miss ratio of 3.7 : 1 (418 windowed measured-drawdown saves against 112 misses). This ratio is a descriptive count, not by itself evidence of filter skill: a fair-game price path also reaches a 50 percent drawdown before a 100 percent rise more often than the reverse, and a rebuilt analysis supersedes it. Correction (October 2026): I withdraw the early-death tier, the wider 14.8 : 1 ratio that rested on it, and the matched lifecycle comparison (48.9 versus 57.6 percent of mints reaching the `gone` state). In 1,235 of the 1,236 single-sample "early-death" events the same token was rejected again about 10 minutes later and the tracker restarted its record, so these events are tracker artefacts, not token deaths, and `gone` is a tracker state, not confirmation of a rug pull. I also withdraw the claim that every adequately sampled filter is individually net-positive (filter_7 has 3 saves and 3 misses). Per-filter estimates may be affected because follow-up records are keyed by token rather than by rejection decision. Details are in the correction notice on page 1.

**Correction notice (October 2026).** This notice corrects the version of 30 May 2026 and agrees with the v2.1 notice of the companion dataset (Zenodo concept DOI 10.5281/zenodo.19987695, 28 September 2026). The data are unchanged; apart from the abstract, the title-block version line, the Conflict of Interest section and the boxed notes, the text below is the earlier text. (1) I withdraw the early-death tier (Tier 3 in Section III.B) and the combined 14.8 : 1 ratio that rested on it. The tracker kept one record per token and restarted it whenever the token was rejected again: in 1,235 of the 1,236 single-sample events the same token was rejected again about 10 minutes later and was priced again under the new event. These events are tracker artefacts, not tokens that disappeared. (2) I withdraw the matched lifecycle comparison of Section V.C (48.9 versus 57.6 percent of mints reaching `gone`) and every conclusion drawn from it, including the statements that the lifecycle data support the 3.7 : 1 ratio. The early-death label marks re-rejection timing, and `gone` is a tracker state that tokens also leave, not confirmation of a rug pull. (3) The 3.7 : 1 ratio (418 saves against 112 misses) is a descriptive count of windowed measured drawdowns. It is not by itself evidence of filter skill, because a fair-game price path also reaches the 50 percent barrier before the 200 percent barrier more often than the reverse; a rebuilt analysis that reconstructs each price path from all samples of the token supersedes it. (4) I withdraw the claim that every active filter with adequate sample size is individually net-positive: filter_7 has 3 saves and 3 misses. Per-filter results may also be affected because the follow-up records are keyed by token, not by rejection decision: where a token had been rejected under a different filter in the previous 24 hours, the log carries the earlier filter's label in 365 of 481 cases. (5) The tracker sampled every 10 minutes, not every fifteen to thirty minutes with thinning (Section III.A); `liquidity` and `volume24h` are US-dollar values, not log2-binned (Section IV); and Figures 1 and 2 illustrate the withdrawn tier. (6) I developed the audited system and have pending U.S. provisional patent applications related to it; the earlier statement of no competing interests is corrected.

## I. Introduction

Filter-gated algorithmic trading systems on decentralised exchanges (DEXs) reject most candidate tokens they evaluate. The systems assume that rejection criteria save capital more often than they forgo profit. They avoid tokens that would have produced losses; they reject tokens that would have produced wins. Whether the assumption holds is testable only against the observed post-rejection outcomes of the specific rejected candidates.

Prior work by the author [Kamat 2026b] introduced Post-Rejection Follow-up Sampling (PRFS). PRFS collects forward price-and-liquidity observations on rejected tokens at configurable intervals, over a follow-up horizon of up to twenty-four hours. The present paper reports the first at-scale precision audit produced by applying a five-tier outcome-classification rule to a PRFS dataset.

The contribution is twofold. First, a methodological refinement: a five-tier classification rule that augments the traditional two-tier (saved / missed) audit with three intermediate tiers (saved-by-early-disappearance, flat, unclassifiable). Second, an empirical result: applied to a 13-day window, the aggregate save-to-miss ratio under the conservative two-tier interpretation is 3.7 : 1. Every active filter with adequate sample size is individually net-positive on that criterion. The wider five-tier audit yields a more favourable 14.8 : 1 ratio. The single-sample-disappearance proxy that drives the additional saves does not validate as an immediate-rug signal against lifecycle ground truth. We therefore report the conservative result as the headline and treat the wider audit as a sensitivity analysis.

## II. Related Work

Reject-side outcome inference is a long-standing problem in econometrics and machine learning. In credit scoring, "reject inference" techniques [Crook and Banasik 2004] estimate the counterfactual outcomes of rejected applicants from observed-accepted-applicant data under selection-bias correction. In algorithmic trading, traditional backtest evaluation [Fang et al. 2022] reconstructs rejected-candidate trajectories from price-feed archives. The MEV literature on transaction-level outcomes [Daian et al. 2020] concerns realised outcomes of executed trades, not counterfactuals of rejected candidates. PRFS and the present audit differ from these prior approaches in that the forward-outcome data is collected in the live market environment in which the rejection occurred, and the classification rule is applied directly to the observed forward trajectory rather than to a reconstructed approximation of it.

## III. Methodology

### III.A The Post-Rejection Follow-up Sampling Substrate

The PRFS methodology is described in detail in [Kamat 2026b]. Briefly, a separate tracking subsystem samples each rejected token's price and liquidity at configurable intervals (in this deployment, every fifteen to thirty minutes during the first hour after rejection, thinning thereafter) for up to twenty-four hours after rejection. Each sample produces a row in `rejection_outcomes` keyed by mint and rejection timestamp; a sampling job terminates on horizon expiry, on disappearance of the token from the price oracle, or on retry-budget exhaustion.

### III.B Five-Tier Classification Rule

We classify each rejection event into one of five outcome tiers:

1. **Saved (windowed):** at least two forward samples within the 24-hour window, with the minimum sampled price reaching or falling below 50 percent of the reference price (the smallest-ageMin sample's price). Interpretation: had the system entered, it would have experienced a measured drawdown of at least 50 percent within 24 hours.

2. **Missed (moon):** at least two forward samples within the 24-hour window, with the maximum sampled price reaching or exceeding 200 percent of the reference price. Interpretation: had the system entered, it would have experienced a 100 percent run-up within 24 hours.

3. **Saved (early-death):** exactly one forward sample within the 24-hour window, collected at an ageMin of 60 minutes or less. **Interpretive premise:** such single-sample events are interpreted as tokens that disappeared from the price oracle before a second sample could be collected, on the working hypothesis that the disappearance signals rug-pull or delisting (outcomes the filter stack is designed to avoid). We report results under this interpretation and caveat it against the external validation in §V.C.

4. **Flat:** at least two forward samples within the 24-hour window, with neither the saved nor missed thresholds reached. Interpretation: the token's forward trajectory was bounded and the rejection neither saved capital nor forwent profit at the threshold levels examined.

5. **Unclassifiable:** invalid reference price (NaN or non-positive), or single-sample event at ageMin > 60 minutes (insufficient evidence to classify).

**Precedence between Tiers 1 and 2 (tie-break).** A single group of forward samples can satisfy both the saved-windowed condition (trough ≤ 50 percent of reference) and the missed-moon condition (peak ≥ 200 percent of reference) within the same 24-hour window, that is, a token can both halve and double during the follow-up. When both conditions hold, the event is classified as **missed**. This precedence (missed-over-saved) matches the deposited audit script's branching order and is the necessary tie-break for the per-filter percentages of §V.A and the conservative 3.7 : 1 ratio of §V.B to reproduce from the rule alone; a reader implementing §III.B with the opposite precedence would obtain a different per-filter table and a higher conservative save-to-miss ratio. The tie-break affects a small minority of events in the deposited cohort.

The audit script implementing this rule is included in the companion deposit (`Kamat_FilterPrecision_2026_audit.js`); the script is single-pass, deterministic, and produces the same classification distribution on any run against the deposited file.

### III.C Aggregate Ratios

From the per-filter tiered counts we derive two aggregate ratios.

The **conservative save-to-miss ratio** uses only the saved-windowed tier in the numerator and the missed-moon tier in the denominator. This ratio rests on measured forward price movements alone and does not depend on any interpretive premise about single-sample events.

The **combined save-to-miss ratio** sums saved-windowed and saved-early-death in the numerator. This ratio is more favourable to the filter stack but depends on the interpretive premise that single-sample-within-60-minute events represent tokens that disappeared from the price oracle for rug-pull or delisting reasons.

We lead with the conservative ratio. The combined ratio is reported as a sensitivity analysis whose interpretive premise is examined against lifecycle ground truth in §V.C.

## IV. Data Sources

The audit operates on the frozen PRFS observation window deposited as the companion dataset (Zenodo DOI 10.5281/zenodo.19987695). The window spans 2026-04-10T21:54Z through 2026-04-23T15:22Z UTC, approximately 13 calendar days (12.7 elapsed days). The v2 deposit contains:

- `Kamat_FilterPrecision_2026_outcomes.jsonl`, 99,510 sample rows on the documented eleven-field schema (`sampleTs`, `mint`, `symbol`, `rejectReason`, `rejectTs`, `ageMin`, `priceUsd`, `liquidity` log2-binned, `volume24h` log2-binned, `dexId`, `pairAddress`)
- `Kamat_FilterPrecision_2026_audit.js`, the reference audit-script implementation
- `Kamat_FilterPrecision_2026_figures.py`, the figure-generation script
- Two figure PNGs (age distribution; sample cadence)
- `PROVENANCE.md` and `README.md`

Filter labels in the deposit are anonymised to `filter_1` through `filter_8` in descending order of count.

The v2 deposit (2026-05-30) supersedes a v1 deposit (2026-05-02) that contained three issues subsequently corrected. First, the v1 JSONL shipped with the live filter-rule names as `rejectReason` field values, conflicting with the manuscript's anonymisation claim; the v2 substitutes `filter_1` through `filter_8` deterministically by descending count. Second, the v1 JSONL contained internal columns the deposit schema documents as removed; the v2 strips every non-documented column. Third, the v1 file's `source` column carried internal collector identifiers; the v2 removes the column entirely so the deposited schema exactly matches the eleven documented fields with no additional optional columns. A small minority of alternate-schema rows in v1 (which the audit script silently ignored because they lacked the audit-required fields) are dropped, reducing the row count from 99,858 to 99,510. The v2 audit script produces the same classification distribution as the v1 audit script: 2,402 groups, 418 saved-windowed, 1,236 saved-early-death, 636 flat, 112 missed, 0 unclassifiable. The deposit README documents the substitution table, the column-strip rule, and a whitelist-based verification snippet that confirms the file's key set equals the documented eleven-field schema exactly.

## V. Results

### V.A Outcome Distribution by Filter

Applying the five-tier rule to the deposited JSONL via the deposited audit script:

| Filter | N | Saved-W % | Saved-ED % | Flat % | Missed % | Unclass % |
|---|---|---|---|---|---|---|
| filter_1 | 935 | 17.9 | 44.8 | 33.7 | 3.6 | 0.0 |
| filter_2 | 572 | 16.6 | 58.9 | 18.9 | 5.6 | 0.0 |
| filter_3 | 361 | 15.2 | 50.1 | 31.0 | 3.6 | 0.0 |
| filter_4 | 215 | 20.0 | 58.6 | 15.8 | 5.6 | 0.0 |
| filter_5 | 210 | 14.8 | 65.7 | 15.7 | 3.8 | 0.0 |
| filter_6 | 80 | 30.0 | 37.5 | 20.0 | 12.5 | 0.0 |
| filter_7 | 25 | 12.0 | 16.0 | 60.0 | 12.0 | 0.0 |
| filter_8 | 4 | 0.0 | 25.0 | 75.0 | 0.0 | 0.0 |
| **Total** | **2,402** | 17.4 | 51.5 | 26.6 | 4.7 | 0.0 |

(The per-filter percentages and the aggregate distribution above reproduce exactly from the deposited audit script against the deposited file. They are unchanged from v1.)

### V.B Aggregate Save-to-Miss Ratios

> **Corrected (October 2026).** The combined 14.8 : 1 ratio and the statement that filters 1 to 7 are each net-positive are withdrawn (filter_7: 3 saves, 3 misses). The 3.7 : 1 ratio is descriptive only. See the correction notice, items 1, 3 and 4.

**Conservative interpretation (saved-windowed only):** 418 saved-windowed events against 112 missed events yields a ratio of **3.7 : 1**. Every individual active filter (filter_1 through filter_7) is net-positive on this criterion (saved-windowed % > missed %); filter_8 has insufficient sample size (n=4) for a confident per-filter verdict but its missed % is zero.

**Combined interpretation (saved-windowed plus saved-early-death):** 1,654 combined saves against 112 missed events yields a ratio of **14.8 : 1**. The interpretive premise of this number, that the 1,236 early-death events represent tokens that disappeared from the price oracle for rug-pull or delisting reasons, is examined in §V.C.

### V.C External Validation of the Early-Death Tier

> **Withdrawn (October 2026).** This matched comparison and every conclusion drawn from it are withdrawn: the early-death label marks re-rejection timing, not token death, and `gone` is a tracker state. The text is kept unchanged for the record. See the correction notice, item 2.

A separately-published dataset by the same author [Kamat 2026d] (RED-2400, Zenodo DOI 10.5281/zenodo.19989075) includes a graveyard-tracker file (`graveyard_lifecycle.csv`) that records subsequent lifecycle observations of rejected tokens. The RED-2400 deposit makes a direct test of the early-death tier's interpretive premise feasible by joining the early-death-classified mints of the present deposit to the graveyard observations of the RED-2400 deposit.

We performed this join. Of the 399 distinct mints classified as `saved_early_death` by the audit script on the present deposit, all 399 have at least one subsequent lifecycle observation in the RED-2400 `graveyard_lifecycle.csv`. Of those 399 mints, **195 (48.9 percent) reach the `gone` state** at some point in the observation window; 204 (51.1 percent) are observed alive in either `alive_active` or `alive_dormant` after the supposed "early death."

The matched comparison group is the other rejected mints with lifecycle observation: tokens that were rejected by the filter stack but did not receive the early-death classification. Of 656 rejected mints with graveyard observation, 343 reach `gone` (52.3 percent overall rejected-mint base rate). Removing the 399 early-death-classified mints leaves 257 non-early-death rejected mints, of which 148 reach `gone`, **a 57.6 percent rate**. The early-death-classified mints reach `gone` at 48.9 percent; the non-early-death rejected mints reach `gone` at 57.6 percent. **The early-death classification therefore identifies tokens that die at a slightly lower rate than other rejected tokens, not a higher rate, in the matched comparison.**

A comparison against the broader graveyard universe (all 1,091 tracked mints regardless of rejection status, 377 reach `gone`, base rate 34.6 percent) would appear to show that the early-death rate is elevated. That comparison is confounded: the filter stack preferentially rejects the lower-quality tokens, so the broader-universe base rate is depressed relative to the rejected subpopulation by design. The matched comparison (rejected versus rejected, with and without the early-death flag) is the correct test for whether the flag carries incremental rug-pull signal beyond rejection itself. It does not.

The implication for the combined ratio is direct: the 1,236 early-death "saves" in the numerator of the 14.8 : 1 figure do not identify a population at elevated immediate rug-pull risk relative to the rejection-only base; the saves they contribute are not differentially supported by the underlying lifecycle ground truth. The aggregate ratio therefore rests on a classification tier whose precision against the matched comparison is no better than rejection alone. The conservative 3.7 : 1 ratio, which does not rely on this tier, is the report supported by the data.

### V.D Headline

> **Corrected (October 2026).** The net-positive statements and the reliance on Section V.C below are withdrawn; the 3.7 : 1 ratio is descriptive only. See the correction notice, items 2 to 4.

The headline finding is the **conservative 3.7 : 1 ratio** under the saved-windowed-only interpretation. The production filter stack is net-positive against an observed-forward-market evaluation: every active filter with adequate sample size is individually net-positive (saved-windowed % > missed % per filter), and the aggregate ratio is 3.7 : 1. The combined 14.8 : 1 figure is reported in §V.B as a sensitivity analysis under an interpretive premise that the matched-comparison test in §V.C does not support.

## VI. Discussion

### VI.A Limitations

The principal limitation of the audit is the one §V.C documents directly: the early-death classification tier's premise is not validated against the lifecycle ground truth available in the related RED-2400 deposit. The conservative 3.7 : 1 result does not depend on this tier, but the wider 14.8 : 1 result does.

A secondary limitation is the observation window length (approximately 13 days). The audit reflects a single market regime; cross-regime stability of the per-filter outcome distribution is not testable from a single 13-day window.

A third limitation is the deployment-specificity of the result. The audit reflects a single production trading-system deployment on a single chain (Solana) at a single DEX surface. The per-filter shares and the precise ratios are deployment-specific; the methodology generalises.

### VI.B Implications for Filter Design

Under the conservative interpretation, the production filter stack is net-positive across the seven active rules with adequate sample size (filter_1 through filter_7); filter_8 has insufficient sample (n=4) for a confident per-filter verdict, although its missed percentage is zero. The per-filter saved-windowed percentages (12.0 % through 30.0 % across the qualifying filters) indicate that the rules vary substantially in how often they save capital under measured drawdown, with filter_6 (n=80) showing the highest saved-windowed share. The per-filter missed percentages (0.0 % through 12.5 %) indicate that the rules also vary in how often they forgo profit; the higher missed percentages co-occur with the smaller-N filters, where statistical noise is also higher. A finer cross-deployment analysis is appropriate follow-up.

## VII. Conclusion

> **Corrected (October 2026).** The net-positive claim, the external validation and the conclusions drawn from the matched comparison are withdrawn; see the correction notice on page 1.

We have reported a precision audit of a production filter stack against a 13-day window of post-rejection follow-up observations. Under the conservative saved-windowed-only interpretation, the aggregate save-to-miss ratio is 3.7 : 1 and every active filter with adequate sample size is net-positive. A wider combined interpretation that additionally credits single-sample-within-60-minute events as saves yields a 14.8 : 1 ratio. An external validation of that wider interpretation against the matched comparison group of other rejected mints in the RED-2400 deposit shows that early-death-classified tokens reach the lifecycle `gone` state at a slightly lower rate (48.9 percent) than non-early-death rejected mints (57.6 percent); the early-death tier therefore does not identify incremental rug-pull risk over rejection itself, and the apparent elevation of the early-death rate against the broader-universe base rate (34.6 percent) is an artifact of filter selection rather than a property of the classification. The methodological contribution is the separation of the two evidence bases and the demonstration that the wider tier does not survive the matched-comparison test. The empirical contribution is the conservative 3.7 : 1 ratio.

---

## Figures

> **Withdrawn (October 2026).** Both figures illustrate the withdrawn early-death tier. Every single-sample event has its sample 0 to 10 minutes after the rejection because the token was rejected again and the tracker restarted its record; the interpretations in the captions are withdrawn. See the correction notice, items 1 and 5.

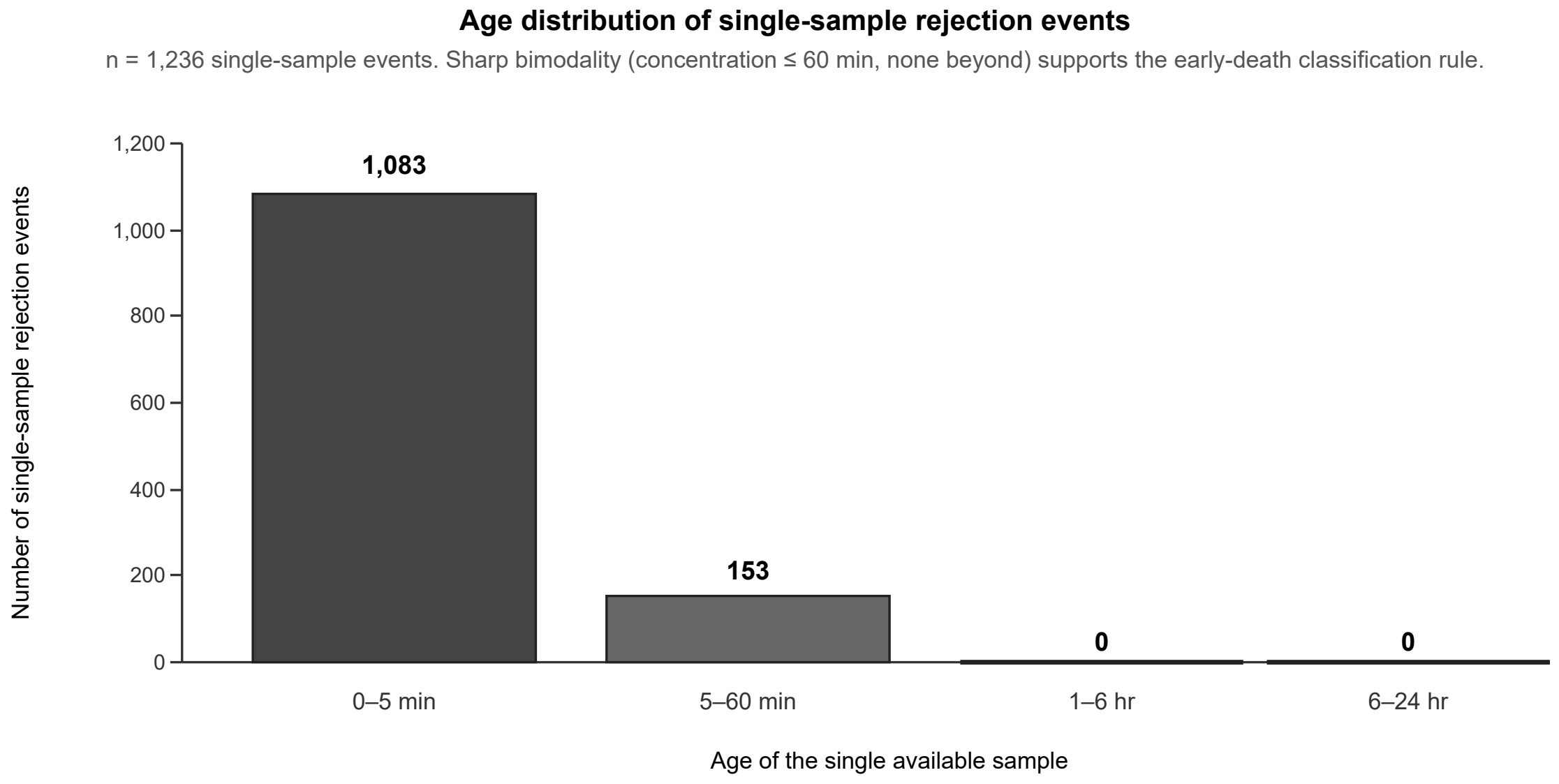

***Figure 1*** *The near-total concentration of single-sample events in the first hour after rejection, and the absence of any single-sample event beyond 60 minutes, motivates and supports the early-death classification rule of §III.B. Counts: 1,083 events in the 0–5 minute bucket; 153 events in the 5–60 minute bucket; zero events in the 1–6 hour and 6–24 hour buckets. The bimodality is qualitatively distinct from the roughly uniform distribution that would arise under random tracker failure or sporadic oracle outage.*

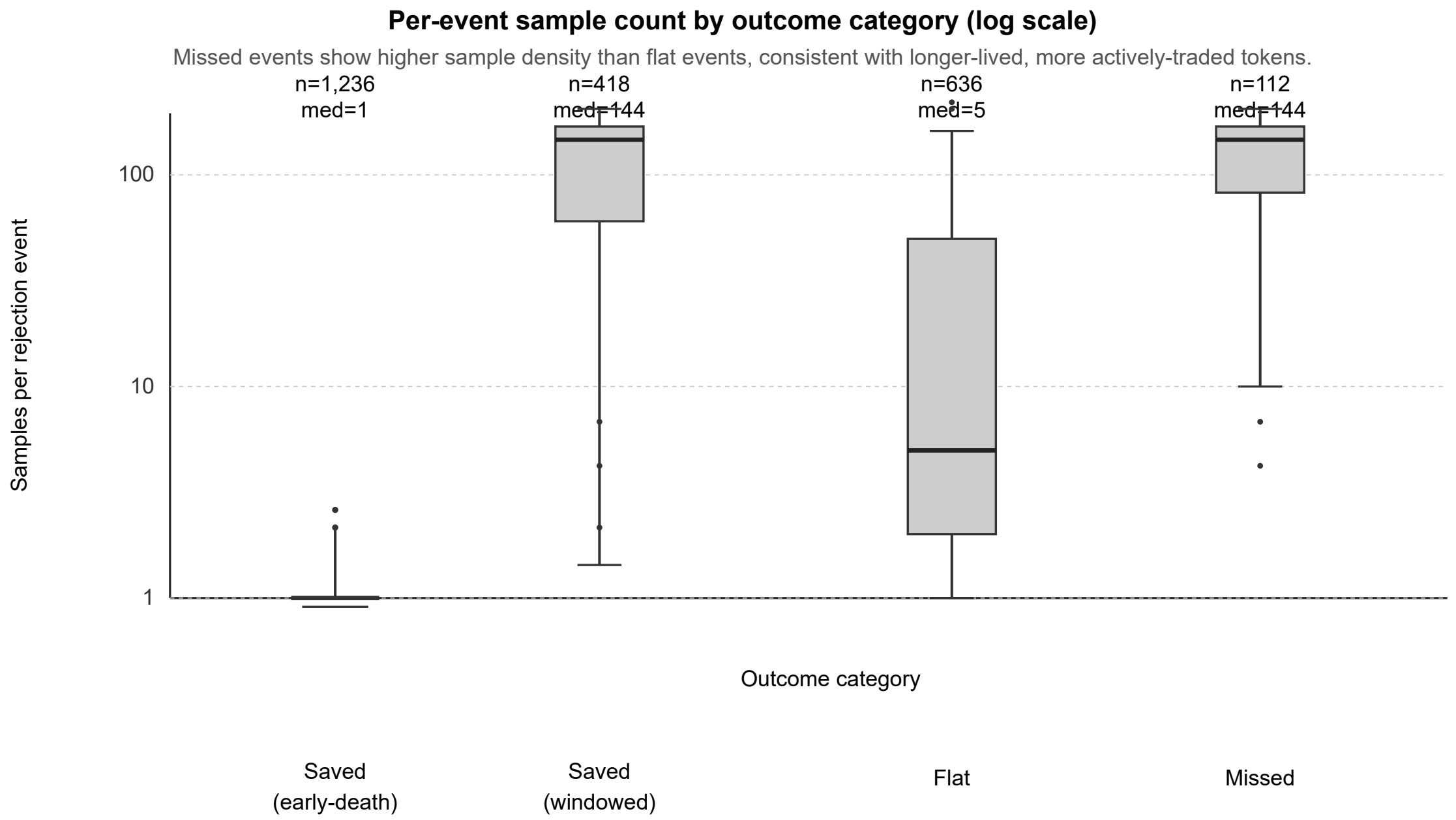

***Figure 2*** *Box plot per outcome tier: Saved (early-death), Saved (windowed), Flat, Missed. Median values: 1, 144, 5, 144 respectively. Sample counts: 1,236; 418; 636; 112. Missed events show systematically higher per-event sample counts than flat events, consistent with the interpretation that missed events correspond to tokens with enough liquidity and activity to remain observable and to generate meaningful price movement, whereas flat events correspond to lower-activity tokens that remain alive without material movement.*

## References

[1] J. Crook and J. Banasik, "Does reject inference really improve the performance of application scoring models?" *Journal of Banking and Finance*, vol. 28, no. 4, pp. 857–874, 2004.

[2] F. Fang, C. Ventre, M. Basios, L. Kanthan, D. Martinez-Rego, F. Wu, and L. Li, "Cryptocurrency trading: A comprehensive survey," *Financial Innovation*, vol. 8, no. 1, pp. 1–59, 2022.

[3] P. Daian, S. Goldfeder, T. Kell, Y. Li, X. Zhao, I. Bentov, L. Breidenbach, and A. Juels, "Flash Boys 2.0: Frontrunning in decentralized exchanges, miner extractable value, and consensus instability," in *Proc. IEEE Symposium on Security and Privacy*, 2020, pp. 910–927.

[4] A. Kamat, *RED-2400: A Public Benchmark of Algorithmically-Rejected Trading Events with Outcome Labels* (v2, 2026-05-30). SSRN abstract_id 6702198; Zenodo DOI 10.5281/zenodo.19989075. [Kamat 2026d]

[5] A. Kamat, *Post-Rejection Follow-up Sampling: A Methodology for Counterfactual Outcome Measurement in Algorithmic DEX Trading* (v2, 2026-05-30). SSRN abstract_id 6607301; Zenodo DOI 10.5281/zenodo.20043516. [Kamat 2026b]

---

## Version History

- **Correction notice (October 2026):** early-death tier, 14.8 : 1 ratio, Section V.C and the net-positive claim withdrawn; competing-interests statement corrected. See page 1.
- **2.0 (2026-05-30):** Version corrected by the notice above. Companion dataset at Zenodo DOI 10.5281/zenodo.19987695.
- **1.0 (2026-04-24):** Superseded.

## Conflict of Interest

Corrected (October 2026): I developed the trading system whose rejections are audited here and have pending U.S. provisional patent applications related to it. The earlier version of this section declared no competing interests.